\documentclass[preprint,showpacs,aps,amsmath,amssymb,pre,preprintnumbers]{revtex4}

\newcommand{\BE}{\begin{equation}}
\newcommand{\EE}{\end{equation}}
\newcommand{\BA}{\begin{eqnarray}}
\newcommand{\EA}{\end{eqnarray}}
\usepackage{graphicx}% Include figure files
\usepackage{dcolumn}% Align table columns on decimal point
\usepackage{bm}% bold math
\input epsf.tex
\preprint{To appear in Chinese J. Physics
(http://psroc.phys.ntu.edu.tw/cjp/index.php)}% cond-mat/0408477}

\begin{document}

\title{About negative refraction and left handed materials}

\author{Zhen Ye}\email{zhen@phy.ncu.edu.tw}
\affiliation{Wave Phenomena Laboratory, Department of Physics,
National Central University, Chungli, Taiwan 32054}

\date{May 4, 2004}

\begin{abstract}

Here I present an overview of recent studies of the phenomenon of
negative refraction and left-handed materials. I will discuss some
basic questions pertinent to the problem. It is pointed out that
the current claims of negative refraction and left handed
materials are not conclusive. To support our consideration, I will
start with the fundamental physics.

\end{abstract}
 \maketitle

\section{Prelude}

Waves surround us. Direct human communication is mainly conveyed
by acoustic waves, and is enriched by gestures which are passed
into our eyes through optical waves. Indirect human conversation
is transmitted via electromagnetic (EM) waves. Nowadays electronic
waves are also everywhere in our daily experiences such as audio
\& video systems, computers, and Nintendo games. Among many
interesting properties pertinent to waves, refraction is probably
the most common phenomenon in our daily life.

In simple terms, refraction refers to waves bending in passing
obliquely from one medium to another. It is a common observation
that a beam of light bends when it enters glass or water at a
tilted angle. Rainbow which must be a phenomenon known to every
one who could see is another example of wave refraction. More
specifically, refraction occurs when waves pass across the
interface between two different effectively uniform media. Briefly
speaking, the effectiveness refers to that inside the medium a
phase vector ($\vec{k}$) can be effectively defined for a given
frequency. In an isotropic medium, for example, a plane wave can
be supported and is described in the form like
$A\exp(i\vec{k}\cdot\vec{r}- i\omega t)$. Therefore to study the
phenomenon of refraction, it is important to verify whether the
media are effective or not.

In the world with which we are so far most familiar, EM waves or
lights propagate according to the rule of right hand. That is, the
electric and magnetic vector fields $\vec{E}$ and $\vec{H}$, and
the phase vector $\vec{k}$ form a righthanded coordinate system.
In other words, the direction $\vec{k}$ is along the direction of
$\vec{E}\times\vec{H}$. Materials in which $\vec{k}$ and
$\vec{E}\times\vec{H}$ are in the same direction are called
isotropic media. The direction of $\vec{E}\times\vec{H}$ is called
the ray or beam direction.

In nature, there is another class of materials. In these
materials, although the direction of $\vec{k}$ may not be exactly
along the direction of $\vec{E}\times\vec{H}$, the angle between
the two is smaller than 90 degree, in the mathematic terms,
$\vec{k}\cdot(\vec{E}\times\vec{H}) > 0$. This class of materials
is called optically anisotropic media. Calcite crystals and
YVO$_4$ bi-crystals are two examples of anisotropic media.

Wave refraction can appear dramatically different when crossing
the interface between either isotropic or anisotropic media. For
example, when an optical wave crosses an interface between two
isotropic media, the refracted wave will always bend towards the
other side of the incidence with regard to the normal of the
interface. The refraction and incident angles obey Snell's
law\cite{Gen}. When the wave passes an interface between an
isotropic medium and an anisotropic medium, however, a phenomenon
called amphoteric refraction can occur. That is, the refracted ray
may bend either towards the same side as the incidence or the
other side of the incidence, depending on the incident angle and
the orientation of the anisotropic medium. In this case, Snell's
law fails\cite{Yariv}. For a uniaxial crystal, for instance, there
are two eigen modes of plane waves: the ordinary and
extra-ordinary modes\cite{Yariv}. The dispersion relation of the
ordinary mode is isotropic and this mode of light behaves as the
light beam in an ordinary material, while the equal frequency
surface or normal surface\cite{Yariv} of the extra-ordinary mode
is ellipsoidal with the optic axis being the axis of revolution.
The allowed wave vectors are those vectors pointing from the
center of the ellipsoid to its surface. For this mode, the energy
velocity is shown to be equal to the group velocity determined by
$\vec{v}_g = \nabla_{\vec{k}}\omega(\vec{k})$\cite{Yariv}.

\begin{figure}[hbt]
\vspace{10pt} \epsfxsize=2in\epsffile{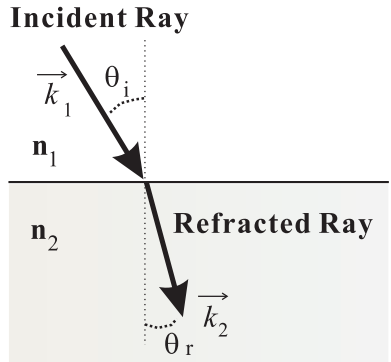}\smallskip
\caption{Illustration of refraction at the interface between two
isotropic media.} \label{fig1}
\end{figure}

Figure~\ref{fig1} is an illustration of refraction across a
boundary between two isotropic media whose refraction indices are
denoted as $n_1$ and $n_2$ respectively. In this case, the ray
direction and the phase vector's direction are the same. The
boundary condition $\vec{k}_1\cdot\vec{r}_\parallel =
\vec{k}_2\cdot\vec{r}_\parallel$ leads to Snell's law \BE
n_1\sin\theta_i = n_2\sin\theta_r,\label{eq:snell}\EE where
$\theta_i$ and $\theta_r$ denote the incidence and refraction
angles respectively.

\begin{figure}[hbt]
\vspace{10pt} \epsfxsize=3.35in\epsffile{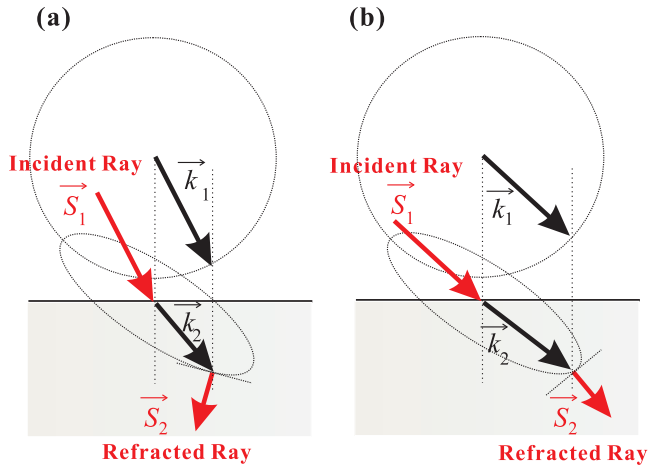}\smallskip
\caption{Illustration of amphoteric refraction at the interface
between an isotropic medium and an anisotropic medium: (a)
negatively appearing refraction; (b) positive refraction}
\label{fig2}
\end{figure}

Figure~\ref{fig2} depicts the amphoteric refraction of the
extraordinary light at a flat interface between an isotropic
medium and an anisotropic medium. The boundary condition requires
that the phase vectors $\vec{k}$ be continuous, when projected
onto the interface. The ray direction, denoted by the time
averaged Poynting vector $\vec{S} =
\frac{1}{2}\mbox{Re}[\vec{E}\times\vec{H}^\star]$, is normal to
the equal frequency surface. Here the refracted ray is bent
towards the same side as the incident ray in (a) while the ray is
tilted to the other side in (b). Such an amphoteric refraction
phenomenon can be readily observed at the interface between air
and calcite crystals\cite{Yau}. In this case, Snell's law in
Eq.~(\ref{eq:snell}) fails. A modified version of the angle
dependence should be introduced\cite{Yau}.

\section{Introduction}

Equipped with the above textbook knowledge, we are ready to
discuss the much debated issue of Left Handed Materials (LHMs) and
the associated phenomenon of negative refraction (NR).

The concept of LHMs was introduced by Veselago many years ago
\cite{Ves}. What is Left Handed Material? According to Veselago
\cite{Ves}, a left handed material refers to electrodynamics of
substance with simultaneously negative values of permittivity
$\epsilon$ and permeability $\mu$. An immediate deduction from
this perception is that the material should have a negative
refraction index, as $n = \sqrt{\epsilon}\sqrt{\mu} = -
\sqrt{|\epsilon\mu|}$. Another important outcome is that the
energy ray $\vec{S}$ and the phase vector $\vec{k}$ will no longer
point to the same direction. Instead, they will be opposite to
each other in an isotropic LHM or make an angle bigger than 90
degree in an anisotropic LHM, i.~e. $\vec{k}\cdot \vec{S}  < 0$.

\begin{figure}[hbt]
\vspace{10pt} \epsfxsize=3.25in\epsffile{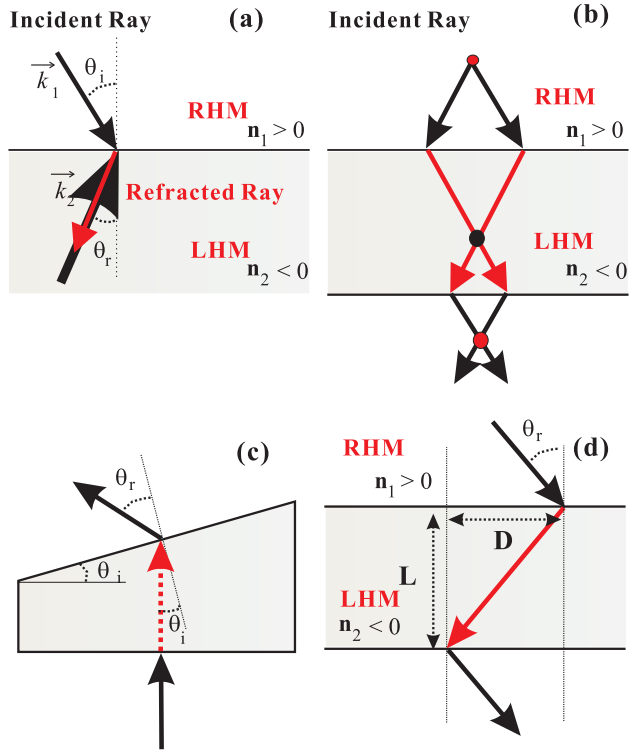}\smallskip
\caption{Illustration of a few expected phenomena for EM wave
propagation through a LHM: (a) negative refraction; (b) Imaging by
a flat slab of LHM; (c) negative refraction by a prism of made of
LHM; (c) negatively shifted outgoing waves. Here the red arrows
denote the energy ray inside the LHM. Here we have ignored the
high order internal reflections.} \label{fig3}
\end{figure}

A negative refractive index would lead to a few peculiar
phenomena. Taking the propagation of EM waves from a normal
material through an isotropic LHM as the example, some of the
expectations are described in Fig.~\ref{fig3}. Inside the LHM, the
phase vector points (Black arrows) to the opposite direction of
the energy rays, denoted by the red arrows. Fig.~\ref{fig3}(a)
illustrates the negative refraction. Here, Snell's law applies
as\BE n_1\sin\theta_i = (-|n_2|)(-|\sin\theta_r|).\EE

The second prediction is the imaging by a slab of LHM as shown in
Fig.~\ref{fig3}(b). For a slab of normal refractive materials,
isotropic or anisotropic, no image can be made across the slab.
There will be no image inside the slab either. Here,
Fig.~\ref{fig3}(b) shows that two images can be made across the
slab of LHM when a wave is transmitted from a point source.

The third possibility is the negative refraction by a prism of
LHMs. This is described in Fig.~\ref{fig3}(c). A plane wave is
incident to the horizontal side of the prism. At the long side,
the incident angle is determined as the prism angle. From the
refraction angle $\theta_r$, one can deduced the refractive index
using Snell's law. If it is tilted towards the left side, the
refractive index is negative. Otherwise, the refractive index is
positive.

When a plane wave is incident on to a flat slab of LHMs, the
outgoing wave will be shifted towards the left side, as shown in
Fig.~\ref{fig3}(d). The negative refractive index can thus be
deduced as \BE n_2 =
-\frac{n_1\sin\theta_i}{\sin(\tan^{-1}(D/L))}. \label{eq:flat}\EE

\begin{figure}[hbt]
\vspace{10pt} \epsfxsize=3.25in\epsffile{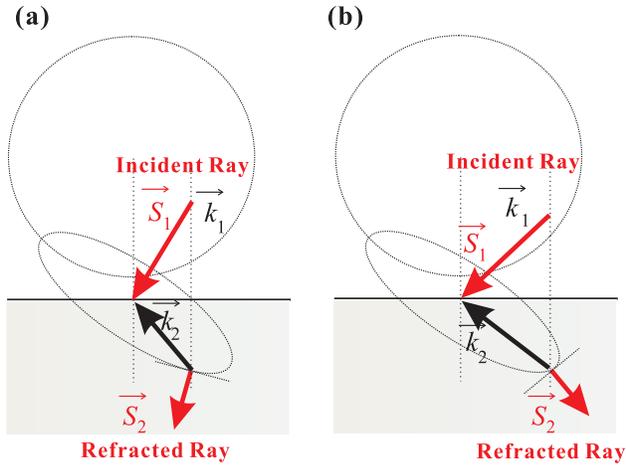}\smallskip
\caption{Illustration of amphoteric refraction of the
extra-ordinary waves at the interface between an isotropic normal
medium and an anisotropic uniaxial LHM medium: (a) positive
refraction; (b) negative refraction. The black and red arrows
denote the phase vectors and energy rays repectively. In isotropic
normal media, the energy ray follows the direction of the phase
vector.} \label{fig4}
\end{figure}

The four deductions described in Fig.~\ref{fig3} have actually
formed the base line in the current searching for LHMs and
negative refraction. A number of notes must be made.
\begin{enumerate}

\item An important component underlying these scenarios is that
the medium can be regarded as an effective medium, in which a
refractive index can be defined, and all the wave propagation will
be fully describable by the Maxwell equations based upon this
index.

\item Another essential point is that the LHM is isotropic, to
warrant the four deductions. For an anisotropic LHM material, the
amphoteric refraction can also appear, as shown in
Fig.~\ref{fig4}. In this case, the only way to differentiate the
amphoteric refraction by a normal anisotropic medium from the
amphoteric refraction by an anisotropic LHM medium is to look at
the phase information. In addition, Snell's law fails for the
anisotropic case; therefore the refractive index cannot be readily
obtained in this case.

\item The negative refraction of LHMs in Fig.~\ref{fig3} must be
differentiated from the negatively appearing refraction due to
anisotropy in Fig.~\ref{fig2}(a). The mechanisms behind the two
similar phenomena are completely different. As shown, the phase
information in the two scenarios is totally different. The
negative refraction of LHMs would be revolutionary not only with
respect to technology, but also to fundamental physics such as
quantum electrodynamic processes. To avoid confusion, we call the
negatively appearing refraction due to anisotropy or any non-LHM
related effect {\bf quasi-negative refraction}, while reserving
the term {\bf negative refraction} for LHMs. Negative refraction
without negative index belongs to the former category. The two
categories are completely different in the mechanism.

\end{enumerate}

The proposal of LHMs by Veselago did not attract much attention
until 2000 when Pendry further pointed out that negative
refraction makes a perfect lens\cite{Pendry}. Ever since then, the
research on such a perfect lens and LHM has been skyrocketing in
the midst of much debate. A great body of literature has been and
continues to be generated.

In spite of a few challenges with regard to the concept of LHM or
relevant negative refraction effects \cite{Garcia1,PE,PE1,YePRE},
the mainstream consensus has been that some indications of
negative refraction effects are affirmative\cite{Pendry2003}:
positively negative. Upon inspection, however, we believe that
previous evidence for negative refraction effects may not be
sufficient. The negatively appearing refraction may not be caused
by negative refraction. In fact, some of the phenomena perhaps
should not have been considered in the context of refraction.
Instead, they may be better characterized such as diffraction,
deflection or the anisotropic scattering phenomenon. In the
present paper, we wish to give a detailed elaboration of this
point of view. Some of present results can be referred to our
unpublished preprints available from the internet.

\section{A brief review of the research on negative refraction}

There are two approaches towards the issue of negative refraction
and LHMs. One is to explore all possible properties or
applications of LHMs. This approach is meaningful if and only if
LHMs indeed exist. The second approach, more difficult but
important and fundamental, is to search for or fabricate LHMs. In
this paper, I will only consider the second approach.

So far, there is only one experimental report about naturally
existing materials for which the negatively appearing refraction
has been vaguely related to the negative refraction\cite{Zhang}.
As discussed later\cite{Yau,Comm}, the negatively appearing
refraction is due to the anisotropy of the materials, and is the
quasi-negative refraction phenomenon as illustrated in
Fig.~\ref{fig2}. It has been pointed out above, it is important to
differentiate the quasi-negative and negative refractions. Failing
to recognize the difference could cause significant
confusion\cite{Report1,Report2}.

In searing for LHMs, two types of artificial materials have been
proposed so far. One type involves metallic wires and split ring
resonators, first made and experimentally tested by Shelby {\it et
al.}\cite{Smith}, later further supported by other groups (e.~g.
Refs.~\cite{Houck,Parazz}). The second type is the two dimensional
photonic crystals made by arrays of parallel cylinders, initiated
theoretically by Notomi\cite{Notomi} and Luo {\it et
al.}\cite{Luo}, later experimentally tested by Cubukcu, {\it et
al.} and Parimi {\it et al.}\cite{Cubukcu,Sridhar,Sridhar2}. Not
withstanding or ignoring other investigations, these explorations
are the basis for later experimental or theoretical adventures in
supporting the negative refraction or LHMs.

All the explorations on LHMs and negative refraction are based on
at least one of the four scenarios shown in Fig.~\ref{fig3}. For
example, we can list as follows: (a) Ref.~\cite{Notomi}; (b)
Refs.~\cite{Houck,Luo,Sridhar}; (c)
Refs.~\cite{Smith,Houck,Parazz}; (d) Ref.~\cite{Cubukcu}. In
particular, the negative refraction index has been obtained for
the resonating wire and ring systems in
Refs.~\cite{Smith,Houck,Parazz} and for photonic crystals in
Ref.~\cite{Cubukcu,Sridhar2}.

\section{Discussion}

\subsection{Foundation of consideration}

The question we wish to discuss is whether the current evidence
for LHMs and the related negative refraction is sufficient or
conclusive. Moreover, if further confirmation is needed, what
could be done next. Our consideration will be based upon the
following logic reasoning.
\begin{enumerate} \item Point I: If a phenomenon has another
possible explanation, then it is insufficient to attribute the
phenomenon to a particular mechanism. In other words, if C appears
to be a common phenomenon which could be due to A and B. Then C
cannot be regarded as the unique consequence of either A or B
alone. It is also insufficient to attribute C to either A or B.
Such a degeneracy has to be verified by finer scrutiny. As
mentioned earlier, measurements at various angles is to isolate
isotropy effects, and the phase information is important to
differentiating the quasi-negative refraction due to anisotropy
and negative refraction by anisotropic LHMs. Like in quantum
mechanics, the angular momentum measurement is important to
differentiate different states at the same energy level when the
conservation law of angular momentum holds. \item Point II: The
evidence should not have any self-conflicting point. \item Point
III: The observation should not be accidental. Further
confirmation should be made to check for robustness, unless the
evidence is sufficient. \item Point IV: The theoretical support
should be independent. That is, the theoretical analysis should be
{\it ab initio}, i.~e. the simulation should not use parameters
from the experiment, unless there are new predictions for
experiments to verify. \item Point V: Experimental verifications
with similar setups and conditions may not be considered as
completely independent.

\end{enumerate}

These points are important. As an illustration, let us consider
the example in Fig.~\ref{fig2}. If only measuring the energy rays
(red arrows) in case (a), then applying Snell's law to the rays,
one may claim that a negative refraction has been observed and a
negative refractive index can be deduced. This declaration is
inappropriate.

First, the claim has to rely on the following assumptions: (1) the
medium is isotropic and can be described as an effective medium;
so Snell's law is applicable; (2) based upon (1), it is expected
that negative refraction can also be observed at other incident
angles even without confirmation. Without verifying the validity
of these two assumptions, one will be unable to exclude the
possibility that the observation is accidental (Point III).

Second, if simply taking the measured data from Fig.~\ref{fig2}(a)
and taking the negative refractive index from applying Snell's
law, one would be able to reproduce the ray direction measured and
may claim as a theoretical support. Clearly, such a theoretical
support is not independent (Point IV). In addition, the support
may work for one particular measurement, but fail in other
extended measurements. For example, an isotropic negative
refractive index would fail to explain the observation in
Fig.~\ref{fig2}(b). Now if one comes back to show that the
refraction is anisotropic, then this will be in conflict with the
presumption (Point II).

Third, the negatively appearing refraction in Fig.~\ref{fig2}(a)
can be explained by the medium anisotropy. Therefore it is
insufficient to declare that the observation is negative
refraction, with the implication of finding a LHM (Point I).

Fourth, different groups may measure the energy rays with
different locations of the detector but with the same setup and
condition, and then claim to reproduce each other's results. This
type of supports is clearly not independent (Point V).

These considerations will guide our following discussion. We must
stress again that we have no reason to criticize the experimental
evidence and theoretical analysis, but we wish to consider whether
they are sufficient or conclusive. If not, what can be done to
improve.

\subsection{Analysis of the evidence and methodology}

\subsubsection{General discussion}

In all the experiments that have been performed so far, only the
wave intensity field, which is proportional to the square modulus
of the EM field, has been measured. There is no report of the
experimental or theoretical analysis of the phase information of
the systems which are claimed to be either LHMs or negatively
refracting. We note that there are theoretical investigations of
the phase behavior of LHMs based on the {\it a priori} assumption
that the systems are LHMs; we will not consider this type of
investigations.

Probing the refraction phenomenon and obtaining refractive index,
without the phase information, will have to rely on the assumption
that the medium must be effective and isotropic. So far, there is
no verification on the isotropy and effectiveness. A few possible
ways to check for the isotropy and effectiveness. The isotropy can
be checked by rotating a sample of the medium. If a medium is
effective and isotropy, the wave propagation inside the medium
will not be sensitive to the fine structures of the medium, and
therefore it will not be changed when the structures are uniformly
disordered. So far, no experiment has been done along this line.
As pointed above, it is improper to use Snell's law to energy ray
paths without confirming the isotropy and effectiveness.

\subsubsection{Negative refraction by composite materials}

The composite materials used in Ref.~\cite{Smith,Houck,Parazz} are
made of periodically arranged metallic wires and and split ring
resonators. The components form a two dimensional square
crystal-lattice. The EM waves are transmitted along the direction
perpendicular to the axis of the wires. The measurements have
adopted the prism scenario described in Fig.~\ref{fig3}(c). The
energy rays are measured and used in Snell's law to obtain the
negative refraction index.

While some questions have already raised in Ref.~\cite{PE,PE1}, a
number of other questions can be raised with this type of
measurements. First, there is no verification for the isotropy of
the media, which could be readily verified by the flat slab
transmission scenario in Fig.~\ref{fig3}(d). Due to their periodic
structured, the materials are unlikely isotropic. If so, it would
not have been proper to invoke Snell's law for the energy rays
(Point II). Second, there is no information about either the wave
propagation inside the media nor about the phase. One possible
solution is to correlate the phase of the outgoing waves with that
of the incident waves. Third, since the composite materials are
made of regularly arranged split ring resonators, possible band
structure effects need to be considered. The essence is to show
that the structured materials can be indeed regarded as an
effective medium in which a plane wave can be established. This
question has already been raised by Pokrovsky and
Efros\cite{PE,PE1}. In addition, the interpretation of the
observation in Ref.~\cite{Smith} has also been questioned for
possible effects due to absorption\cite{Garcia1}.

In Ref.~\cite{Houck}, two prisms of the composite materials are
used. The energy rays are measure and used to deduce the
refraction index. The deviation between the two cases amounts to
at least 40 percent. Reference \cite{Parazz} duplicates the
experiment in Ref.~\cite{Smith} (Point V). The numerical
simulation is not independent (Point IV). As pointed out in
Ref.~\cite{PE1}, a correct computer simulation starting with the
microscopic Maxwell's equation would reproduce the experiments,
but whether the observations can be explained in terms of a
macroscopic Maxwell's equation with both negative permittivity and
permeability remains an open question. Such an {\it ab initio}
simulation is yet to come. In addition, as Pokrovsky and Efros
pointed out\cite{PE1}, there are no doubts on the the results, but
there is a need for solid confirmation of the data interpretation.

\subsubsection{Negative refraction by photonic crystals}

The first proposals of negative refraction by photonic crystals
were from Notomi\cite{Notomi} and Luo {\it et al.}\cite{Luo},
later experimentally tested by Cubukcu, {\it et al.} and Parimi
{\it et al.}\cite{Cubukcu,Sridhar,Sridhar2}.

Most of photonic crystals studied so far are two dimensional
arrays of dielectric cylinders. An advantage of this type of
systems is that it can be solved {\it exactly} by the multiple
scattering theory (MST), first formulated by
Twersky\cite{Twersky}, later reformulated for EM
systems\cite{JOSA,Bikash}, acoustic systems\cite{ChenPRL}, and
water waves\cite{Yue}. Therefore we can have a rigorous
examination or new predictions of this type of photonic crystals.
We will adopt MST.

{\it Imaging inside a photonic crystal.} By the FDTD simulation,
Notomi presented an image focusing inside a photonic crystal made
of drilled holes in a uniform dielectric material. An focused
imaging is presented in Fig.~13 of Ref.~\cite{Notomi}. We have
tried to reproduce the results, but failed, partially because the
parameters have not been clearly given in Ref.~\cite{Notomi}. In
the simulation, however, we found that the snapshots of the
imaging field depend significantly on the simulation time. When
the stable stage has not been reached, some focused images like in
\cite{Notomi} could be exhibited\cite{Tang}.

\begin{figure}[hbt]
\vspace{10pt} \epsfxsize=3.25in\epsffile{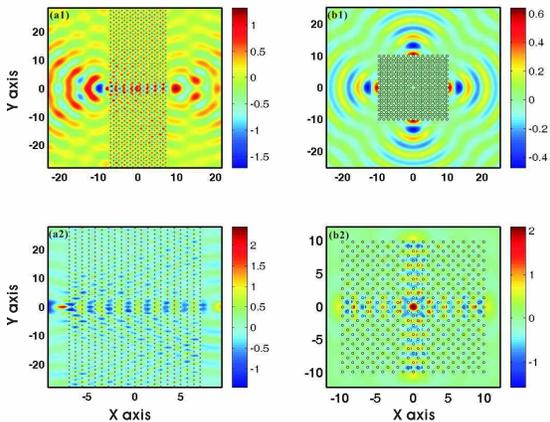}\smallskip
\caption{(a1) and (a2): The imaging fields for a slab of photonic
crystal structure. The slab measures as
10$\sqrt{2}\times40\sqrt{2}$. (b1) and (b2): The imaging fields
for a transmitting source located inside a square array of
cylinders. The square measures $14\sqrt{2}\times 14\sqrt{2}$.}
\label{fig5}
\end{figure}

{\it Flat lens imaging - theoretical.} The flat lens imaging
illustrated by Fig.~\ref{fig3}(b) was first studied in
Ref.~\cite{Luo}. The systems considered are two types of photonic
crystals. One is made of square arrays of dielectric cylinders in
air. The other is made of square arrays drilled holes in a
dielectric medium. Based upon the equal frequency band structure,
the authors show there is a range of frequencies in which an all
angle negative refraction can be archived in this scenario. We
have done two verifications. First, we have reproduced nicely the
observation of Ref.~\cite{Luo}. The results are shown in Fig.~1(c)
of Ref.~\cite{Kuo2}. We indeed observe a focused image across the
flat slab of photonic crystals.

If such an imaging property is due to negative refraction, we
would then have two expectations (Point III): one is to find a
focused image inside the slab as shown in Fig.~\ref{fig3}(b); the
other is to see the negative refraction effect in the scenario
shown in Fig.~\ref{fig3}(d).

For the first expectation, the inside focusing is hard to see in
the arrays of Ref.~\cite{Luo}, as their photonic crystal slab is
too thin. We have thus enlarged the slab and plot both the
amplitude fields and the intensity fields inside the slab. There
is no focused image inside the slab. Instead, we observe that the
energy transmission is guided through the slab, making the image.
This has be further verified by putting an EM wave source inside a
square shaped array of cylinders. The two sets of simulation
results, taken from Ref.~\cite{Kuo2}, are shown in
Fig.~\ref{fig5}. All the parameters and lattice arrangements are
taken from Ref.~\cite{Luo}; all lengths are scaled by the lattice
constant and the frequency is scaled by the radius of the
cylinders. Here we see clearly the image focused on the other side
of the slab in (a) and the four images made at the four exiting
points are due to the guided or collimated wave transmission. It
appears that there are favorable directions for waves to
transport, as shown by (b2). We found that such collimated or
guided transmission is caused by the presence of partial bandgaps.
As can be seen from Fig.~3 of Ref.~\cite{Luo}, the frequency
chosen lies within a partial bandgap in which the waves are
forbidden to transmit along the direction of $\Gamma-X$ in the
square lattice. Due to this partial bandgap, waves seem to avoid
the $\Gamma-X$ direction as much as possible, and thus appear to
be collimated along the $\Gamma M$ direction. The energy paths
shown in Fig.~\ref{fig5} are actually along the $\Gamma-M$
direction.

We have also checked the second expectation. The results are
presented in Ref.~\cite{Chien}. We found that the wave
transmission always tends to be guided or collimated into the
$\Gamma M$ direction, independent of the incident angles or the
orientation of the $\Gamma-M$ direction. This phenomenon cannot be
classified as refraction either in the context of isotropy or
anisotropy.

The results from Refs.~\cite{Kuo2,Chien} led us to propose that
partial bandgaps can serve as a guiding tunnel for the wave
transport, allowing for new applications of photonic
crystals\cite{Chen}. Such a consideration has also been extended
to acoustic systems\cite{Chen5}.

The flat slab imaging discussed in Ref.~\cite{Luo} is not caused
by negative refraction, rather it is caused by directed
diffraction. The misinterpretation in Ref.~\cite{Luo} has also
been pointed out by Li {\it et al.}\cite{Li}. At least, the
explanation in Ref.~\cite{Luo} is unable to exclude other possible
interpretations (Point I).

One may ask for the reason for the discrepancies between the
present results and the analysis in Ref.~\cite{Luo}. Here we would
like to share our thoughts. The problem with the conjectured flat
slab imaging may lie in the approach to the energy flow inside the
slab. The usual approach mainly relies on the curvatures of
frequency bands to infer the energy flow.

As documented in Ref.~\cite{Yariv}, an energy velocity is defined
as $\vec{v}_e = \frac{\frac{1}{V}\int \vec{J}_{\vec{K}}
d^3{r}}{\frac{1}{V}\int U_{\vec{K}} d^3{r}},$ where
$\vec{J}_{\vec{K}}$ and $U_{\vec{K}}$ are the energy flux and
energy density of the eigenmodes, and the integration is performed
in a unit cell, representing a spatially averaged value. It can be
shown that thus defined energy velocity equals the group velocity
obtained as $\vec{v}_g = \nabla_{\vec{K}}\omega(\vec{K}).$
Therefore it is common to calculate the group velocity to infer
the energy velocity and subsequently the energy flows or
refraction of waves. Whether the net current flow through a unit
cell really follows the direction of $\vec{v}_e$ remains unclear.
We note here that the average flux through a surface may be
defined as $\langle\vec{J}\rangle = \frac{\hat{n}}{S}\int
d\vec{S}\cdot \vec{J}$, where $\hat{n}$ is the unit normal vector
of the surface $S$. Clearly, the volume averaged current within a
unit cell does not necessarily correspond to actual energy
flows\cite{GV,PLA}. General speaking, the periodically structured
materials may not be regarded as an effective medium.

The curvature approach fails in the current partial bandgap case.
For example, in the present case, the curvature method would lead
to an almost 80 degree of energy deflection with reference to the
$\Gamma X$ direction when the incidence is at about 25 degree; The
rigorous computational results show that the energy flow is nearly
in the $\Gamma M$ direction\cite{Chien}.

\begin{figure}[hbt]
\vspace{10pt} \epsfxsize=3.25in\epsffile{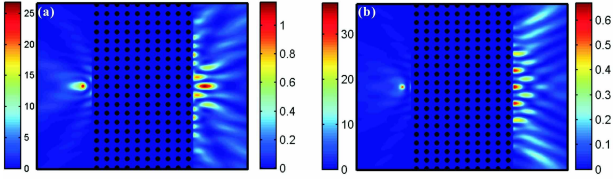}\smallskip
\caption{The image of the intensity-fields for flat slabs. (a) and
(b) refer to two arrangements. In (a), all the arrangements
including the source, the field resolution, and the cylinders are
identical to the experiment\cite{Sridhar}. (b) is the same as (a)
except that the source is move upward by a half lattice constant.
The scales for the fields for the left and right sides are shown
in the figure.} \label{fig6}
\end{figure}

{\it Flat lens imaging - experimental.} Later, flat slab imaging
has been experimentally observed by Parimi et al.\cite{Sridhar}.
The authors reported that an image can be formed cross a flat slab
of photonic crystals made of dielectric cylinders. The appearance
of the image is regarded as the occurrence of negative refraction
and has thus been related to the negative refraction of LHMs. We
found that their interpretation needs to be reconsidered. The
focused image is accidental (Point III). The imaging pattern will
completely change as the source is moved only by a half lattice
constant. From the transmission calculation, we found that the
image is due to the anisotropic scattering of the array of
cylinders. Scattering by a group of materials certainly will yield
bright-dark patterns, simply due to multiple scattering and wave
interference, like in the X-ray imaging. Our rejected comments on
the experimental observation are in Ref.~\cite{Kuo7}. Partial
results are shown in Fig.~\ref{fig6}. Fig.~\ref{fig6}(a)
reproduces remarkably well the experiment. In the case (b), the
source is move upward by a half lattice constant. Here we see that
the imaging field across the slab changes completely. Again, we do
not criticize the experiment, but we think that the interpretation
is ambiguous.

What is the difference between the flat slab imaging due to
partial gaps and the imaging due to anisotropic scattering? The
difference is that the imaging by partial gaps is a stable
phenomenon, while the imaging due to scattering is unstable. The
focusing phenomenon can be even seen by scattering by a single
object under certain manipulation, as long as the scattering is
anisotropic.

{\it Plane wave through a slab of photonic crystals.} Cubukcu {\it
et al.} reported the measurement of a beam of EM waves incident on
 a flat slab of photonic crystals\cite{Cubukcu}, referring to
 Fig.~\ref{fig3}(d). The authors measured the energy ray directions. Then
 Equation (\ref{eq:flat}) was used to obtain
 the refractive index. The author shows that at certain ranges of
 incident angles, the index is negative. As discussed earlier, Snell's law in
 the form of Eq.~(\ref{eq:flat}) is only applicable when the medium
 isotropic. However, the results in Ref.~\cite{Cubukcu} indicate
 the opposite. This is clearly self-conflicting (Point II). In
 addition, we tried to reproduce the results in Fig.~2 of
 Ref.~\cite{Cubukcu}, but failed. We found the results similar to that
 in Ref.~\cite{Chien}. That is, independent of the incident
 angles, the waves are bent towards a particular direction, which
 is across the slab of photonic crystals used in the experiment. A
 possible reason for the discrepancy may be due to the difference
 between the FDTD method used in \cite{Cubukcu} and the MST used
 in our simulation. As mentioned earlier, it takes a considerable long
 time for FDTD results to reach the stable
 stage. Without checking the stability, the FDTD results may be
 unreliable. Another possible cause is the finite size effect.
 In our simulation, we have excluded any thinkable
 numerical artifacts, with regards to the mode numbers, finite number of scatterers,
 boundary effects, and so on.
 We welcome any request for the
 numerical codes and our simulation procedures for independent and unbiased verification.

{\it Prism scenario.} Lastly but not least, we address the prism
 scenario. It is obvious that this scenario can only work when the
 medium is isotropic. Without first verifying that the medium is
 isotropic, the deduction of the refraction index is not
 reliable, as the waves may be already bent at the entering interface. As an example, we
 have simulated two prisms of the photonic crystals made of dielectric cylinders, given in
 Ref.~\cite{Luo,Kuo2}. The results are presented in
 Ref.~\cite{Kuo3}. Here we show partial results.

\begin{figure}[hbt]
\vspace{10pt} \epsfxsize=3.25in\epsffile{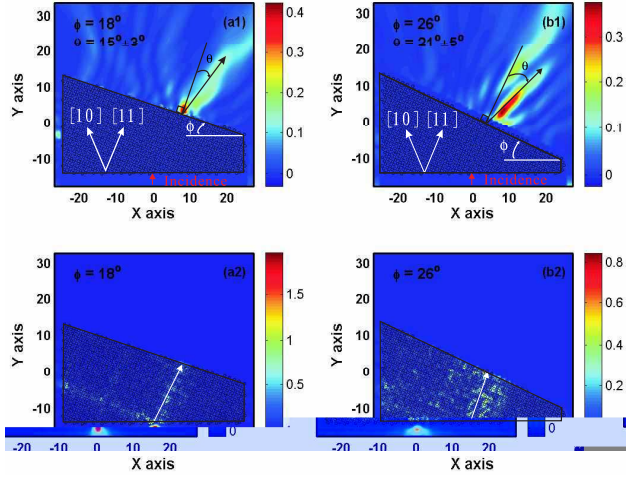}\smallskip
\caption{ The images of the transmitted intensity fields. Here,
the intensities inside and outside the prism structures are
plotted separately, so that the images can be clearly shown with
proper scales. The geometric measurements can be inferred from the
figure. The tilt angles for the two prisms are 18$^o$ and 26$^o$
respectively, and have been labelled in the figures. For cases
(a1) and (b1), we observe the apparent `negative refraction' at
the angles of 15$\pm 3^o$ and $21\pm 5^o$ respectively. When
applying Snell's law, these numbers give rise to the negative
refraction indices of $-0.84 \pm 0.17$ and $-0.87\pm 0.21$ for
(a1) and (b1) separately. In (a2) and (b2), the intensity fields
inside inside the prisms are plotted. Here we clearly see that the
transmission has been bent. In the plots, [10] and [11] denote the
axial directions of the square lattice of the cylinder arrays,
i.~e. $\Gamma-X$ and $\Gamma-M$.} \label{fig7}
\end{figure}

We have plotted the transmitted intensity fields. The results are
presented in Fig.~\ref{fig7}. Here the incident waves are
transmitted vertically from the bottom. The impinging frequency is
0.192$*2\pi c/a$ with $a$ being the radius of the cylinder; the
frequency has been scaled to be non-dimensional in the same way as
in Ref.~\cite{Luo}. We note that at this frequency, the wave
length is about five times of the lattice constant, nearly the
same as that used in the experimental of \cite{Houck}. The
incidence is along the [$\cos 22.5^o$, $\sin 67.5^o$] direction,
that is, the incidence makes an equal angle of 22.5$^o$ with
regard to the [10] and [11] directions of the square lattice; the
reason why we choose this direction will become clear from later
discussions. Moreover, on purpose, the intensity imaging is
plotted for the fields inside and outside the prisms on separate
graphs which have been scripted by `1' and `2' respectively.

First, we ignore the transmitted fields inside the prisms. From
Fig.~\ref{fig7} (a1) and (b1) we are able to calculate the main
paths of the transmitted intensities. The geometries of the
transmission are indicated in the diagrams. The tilt angles of the
prisms are denoted by $\phi$, whereas the angles made by the
outgoing intensities relative to the normals of the titled
interfaces are represented by $\theta$. According to the
prescription outlined in Ref.~\cite{Smith,Houck}, once $\phi$ and
$\theta$ are determined, Snell's law is applied to determine the
effective refractive index: $n\sin\phi = \sin\theta$. If the angle
$\theta$ is towards the higher side of the prisms with reference
to the normal, the angle is considered positive. Otherwise, it is
regarded as negative. In light of these considerations, the
apparent `negative refractions', similar to the experimental
observations\cite{Smith,Houck}, indeed appear and are indicated by
the black arrows in the graphs. After invoking Snell's law, the
negative refraction results are deduced. From (a1) and (b1), we
obtain the negative refractive indices as $-0.84\pm 0.17$ and
$-0.87\pm 0.21$ for the two prisms respectively. The inconsistency
between the two values is less than 4\%. The overall uncertainty
in the present simulation is less than 24\%.  Therefore, a
consistent negative refractive index may be claimed from the
measurements shown in Fig.~\ref{fig7} (a1) and (b1). Furthermore,
we found that with certain adjustments such as rotating the arrays
or varying the filling factor, the index obtained by Snell's law
can be close to the perfect -1.

Now we take into consideration the wave propagation inside the
prisms. As shown by Fig.~\ref{fig1} (a2) and (b2), the
transmission inside the prisms has been bent at the incidence
interfaces, referring to the two white arrowed lines in (a2) and
(b2). If Snell's law were valid at the outgoing interfaces, it
should also be applicable at the incident boundaries. Then with
the zero incidence angle and a finite refraction angle, Snell's
law would lead to an infinite refractive index for the surrounding
medium. Besides, when taking into account the bending inside the
prisms, the incident angle at the outgoing or the tilted surface
is not $\theta$ any more. Therefore the index value obtained above
is incorrect. Our opinion is that we need to differentiate the
refraction from diffraction. The phenomenon shown in
Fig.~\ref{fig7} should be attributed as occurrence of diffraction
effects.

The measurements in Ref.~\cite{Sridhar2} have adopted the prism
scenario. The photonic crystals are made of cylinders, and made
into a prism. Then the energy rays are measured from the incident
and outgoing waves, to obtain refractive index by Snell's law.
Again, the authors have not verified the isotropy. How waves
propagate inside the prism has not been probed. According to the
results in Fig.~\ref{fig7}, the experiment is not sufficient to
identify the LHM behavior of the photonic crystals.

\section{Summary}

In summary, we here discuss some fundamental questions about LHMs
and the associated negative refraction. It is pointed out that the
current experimental evidence is not sufficient to conclude that
LHMs have been formulated. Some concerns and considerations are
presented to support the analysis. The key is to carry out a
comprehensive experimental exploration, with regard to all aspects
of LHMs and negative refraction, including such issues as phase
information, isotropy, and effectiveness etc. The phase
information will be essential to differentiate the negative
refraction from the quasi-negative refraction.

Before talking about either the positive or negative refraction, a
prior task is to find out whether it is the refraction phenomenon
that occurs when waves pass across the interface between two
media. It is insufficient to just look at the deflection of the
energy intensity path to discern the refraction phenomenon.
Otherwise, confusions can accumulate\cite{Report3}. In sonic
crystals, it can be shown that negative refraction can never occur
in the sense of LHMs. We think that the report in
Ref.~\cite{Report3} misleading. Structured materials can certainly
manipulate wave propagation in various ways. However exciting, it
is crucial to confirm first the manipulation is due to the
refraction mechanism before classifying as negative or positive
refraction, and claiming the finding or fabricating of LHMs.

The essence of the present work is not to discard the previous
efforts, rather it is hoped to help clarify some of possible
uncertainties in the previous data interpretation or theoretical
analysis. Only through the detailed examination, debate or
introspection, the understanding of the problem can be deepened.
Given the situation that the issue of LHMs has ignited such a huge
response, a careful reconsideration of the problem is always
something important.

{\bf Acknowldegments} I am grateful to Ting-Kuo Lee, Ding-Ping
Tsai, Han-Fei Yau, Yong Zhang, Ben-Yuan Gu, Zhi-Yuan Li,
Chao-Hsien Kuo, Liang-Shan Chen, M. Nieto-Vesperinas, Dezhang Chu
and many others for constructive correspondences and criticisms.
The help from C.-H. Kuo is particularly thanked. The work received
support from NSC, and the College of Science at NCU.

\end{document}